\documentclass[aps,pre,twocolumn,groupedaddress,amsmath,amssymb,aps]{revtex4-2}
\usepackage{graphicx}% Include figure files
\usepackage{dcolumn}% Align table columns on decimal point
\usepackage{subcaption}
\usepackage{bm}% bold math
\usepackage{xcolor}
\usepackage{soul}
\begin{document}

\title{Energy cascade and Burgers turbulence in the Fermi-Pasta-Ulam-Tsingou chain}% Force line breaks with \\
%\thanks{Partially supported by MIUR PRIN ***.}

\author{Matteo Gallone}
\email{mgallone@sissa.it}
\affiliation{Scuola Internazionale di Studi Superiori Avanzati, Via Bonomea 265, 34136 Trieste, Italy} %\\This line break forced with \textbackslash\textbackslash
%
%\author{Matteo Marian}%
% \email{matteo.marian@studenti.units.it}
%\affiliation{ 
%Department of Physics, University of Trieste, Via A. Valerio 2, 34127 Trieste, Italy} %\\This line break forced with \textbackslash\textbackslash
%%

\author{Antonio Ponno}
 \email{ponno@math.unipd.it}
\affiliation{Department of Mathematics ``T. Levi-Civita'', University of Padova, Via Trieste 63, 35121 Padova, Italy}%\\This line break forced% with \\

\author{Stefano Ruffo}%
 \email{ruffo@sissa.it}
\affiliation{Scuola Internazionale di Studi Superiori Avanzati, Via Bonomea 265, 34136 Trieste, Italy \\
ISC-CNR, via Madonna del Piano 10, 50019 Sesto Fiorentino, Firenze, Italy}%\\This line break forced with \textbackslash\textbackslash

\date{\today}% It is always \today, today,

\begin{abstract}
\noindent

The dynamics of initial long--wavelength excitations of the Fermi-Pasta-Ulam-Tsingou chain has been the subject of intense investigations since the pioneering work of Fermi and collaborators. We have recently found a regime where the spectrum of the Fourier modes decays with a power--law and we have interpreted this regime as a transient turbulence associated with the Burgers equation. In this paper we present the full derivation of the latter equation from the lattice dynamics using an infinite--dimensional Hamiltonian perturbation theory. This theory allows us to relate the time evolution of the Fourier spectrum $E_k$ of the Burgers equation to that of the Fermi-Pasta-Ulam-Tsingou (FPUT) chain. As a consequence, we derive analytically both the shock time and the power--law $-8/3$ of the spectrum at this time. Using the shock time as a unit, we follow numerically the time--evolution of the spectrum and observe the persistence of the power $-2$ over an extensive time window. The exponent $-2$ has been widely discussed in the literature on the Burgers equation. The analysis of the Burgers equation in Fourier space also gives information on the time evolution of the energy of each single mode which, at short time, is also a power--law depending on the $k$-th wavenumber $E_k \sim t^{2k-2}$. This approach to the FPUT dynamics opens the way to a wider study of scaling regimes arising from more general initial conditions.
\end{abstract}

%Use showkeys class option if keyword
\maketitle

\noindent

\section{Introduction}
Understanding the dynamical route to thermalization of physical systems is a fundamental problem in statistical mechanics, since its foundations \cite{Kinchin}. 

At variance with gases, which are characterized by a universal mechanism leading to equilibrium \cite{CerciBook}, solids undergo a more complex relaxation process \cite{Ashcroft-Mermin}, which has not yet been fully understood. This problem is known since the seminal work of Fermi \emph{et al.} \cite{FPU55}, who studied numerically a simple one-dimensional model of a nonlinear crystal. Surprisingly, instead of the expected fast trend to thermal equilibrium, they observed a recurrent, quasi-periodic behaviour. The discrepancy between the expectation of a fast thermalization and the observation of recurrence is named, after the authors, the FPUT paradox (the T being added lately to the acronym in order to acknowledge the significant contribution of Tsingou \cite{Daux08}). It has been realized later that slow relaxation to equilibrium is present for a much larger class of systems. In various cases, the vicinity of the system to an integrable model justifies this phenomenon \cite{Gallavotti-FPU-Libro,Chaos-VolumeCollettivo}. The presence of slow relaxation is not confined to systems of classical mechanics. In the context of quantum mechanics, the interest in the study of thermalization has grown significantly in recent years thanks to the realization of cold atom experiments which produce for the first time the settings studied in classical mechanics \cite{Kino}. Several theoretical interpretations have been proposed, including the theory of generalized Gibbs ensambles \cite{Rigol-PRL}. However, a general understanding of thermalization in quantum mechanics has not yet been achieved and the research field is extremely active. These studies have been extended to quantum systems subjected to external drivings. In this context, slow relaxation to equilibrium is the fundamental process underlying the achievement of Floquet and quasi-Floquet systems \cite{RevFloq}, as theoretically proven in refs. \cite{Abanin,Else,GL}.

\begin{figure}[h!]
	\includegraphics[width=\columnwidth]{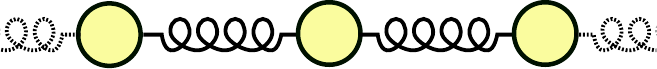}
	
	\caption{Pictorial representation of the FPUT system. The balls are the masses and the springs are nonlinear.}\label{fig:FPUS}
\end{figure}

The physical system introduced by Fermi and collaborators consists of $N$ unit masses connected by non--linear springs (see Fig.\ \ref{fig:FPUS}). The Hamiltonian of the model is
\begin{equation}\label{eq:H}
	H=\sum_{j \in \mathbb{Z}_N} \Bigg[\frac{p_j^2}{2}+V(q_{j+1}-q_j) \Bigg] \, ,
\end{equation}
where $\mathbb{Z}_N=\{1,\dots,N\}$ encoding the periodic boundary conditions and
\begin{equation}
	V(z)=\frac{z^2}{2}+\alpha \frac{z^3}{3}+\beta\frac{z^4}{4} \, .
\end{equation}
If the non--linear part of the interaction vanishes, the dynamics of the \emph{Fourier Energy Spectrum (FES)} becomes trivial because there is no energy exchange among the normal modes of the linearized system. Non--linearity, on the other hand, couples the modes during their time evolution and enables an exchange of energy, which is the mechanism driving thermalization. However, the situation is more nuanced than this, as mode--coupling also occurs in non--linear integrable systems like the Toda chain \cite{TodaIntegrable}, where thermalization, in the Gibbs ensemble sense, does not take place. In fact, at small specific energies $\varepsilon=E/N$, the FPUT model can be seen as a perturbation of the integrable Toda chain. Regardless of the initial condition, the observed phenomena at the nonlinear level resemble those of an integrable system.

In the regime of small specific energies, when a long wavelength initial state is excited, the nonlinearity is able to effectively induce an exchange of energy only among a few modes, resulting in a FES which decays exponentially in the mode number $k$ as $E_k \sim e^{-A|k|/(N \varepsilon^{1/4})}$, on a time--scale increasing as an inverse power--law of $\varepsilon$ \cite{Po03}. In this energy range, the FES has a complex time evolution characterized by a slow energy transfer to higher wavelengths which finally leads to equipartition at times $t \sim \varepsilon^{-9/4}$ (see \cite{BP11}).

At high specific energies, Izrailev and Chirikov found that thermalization occurs on a much faster time--scale \cite{IC66}. The presence of fast thermalization was guessed analytically using the resonance overlap criterion and confirmed in early numerical experiments \cite{Izrailev2}. A thorough numerical work was devoted to the investigation of the intermediate energy region between fast and slow thermalization. It was finally found that this region is intensive in the specific energy \cite{PRARuffo}. When the energy is above this region, the system is also found to be fully chaotic and to support a completely positive Lyapunov spectrum \cite{JPA-Livi-Politi-Ruffo}. 

Recently, in \cite{PRL}, we have analyzed a regime where the specific energy $\varepsilon$ is sufficiently large to allow energy exchange across a wide range of mode numbers. However, the specific energy is still low enough to hinder rapid thermalization of the system. In this regime, for long wavelength initial conditions, the dynamics can be effectively described by a (generalized) inviscid Burgers equation \cite{FF}. This description allows for the prediction of an extensive window of long wavelength Fourier modes where the FES decays with the time-dependent power--law
\begin{equation}\label{eq:DefZeta}
E_k \sim k^{-\zeta(t)}\ \ ;\ \ k_0\leq k\leq k_c\ ,
\end{equation}
with $k_0$ and $k_c$ slowly varying over time. This window sets in at a specific time $t_s$, which corresponds to the shock time of the generalized Burgers equation. We have identified $t_s$ as a natural time--scale for the dynamics of the FES towards equipartition, encoding the parameters of both the model and the initial conditions. Consequently, it is physically convenient to describe the time--evolution to equipartition in units of $t_s$. The power--law exponent at the shock time is $\zeta(t_s) = 8/3$ and it decreases to approximately $2$ around $2t_s$. Beyond this time, it remains constant while the window width, $k_c-k_0$, gradually decreases over time. This behaviour is reminiscent of fluid turbulence, featuring an inertial range $[k_0, k_c]$ over which the FES follows a power--law decay. However, in this context, due to the absence of energy injection and dissipation, we observe a transient turbulence phenomenon. Similarly to fluid turbulence, an exponential decay of the FES occurs beyond the inertial range, i.e., for $k > k_c$. In fluids, this is attributed to a small-scale balance between nonlinearity and dissipation, whereas in our case, dispersion acts as the dissipation mechanism.

As we have mentioned, the FPUT system remains close to the integrable Toda chain for short times and small enough energy \cite{Man}. It turns out that the time evolution of the Toda integrals of motion along the FPUT trajectories separate exactly at the Burgers shock time \cite{Hofstrand}. This result confirms on one hand the key role of the shock time in the process of equipartition, on the other hand does not imply that Burgers equation cannot describe the time evolution even after the shock time.

The discovery of a transient Burgers turbulent regime in the FPUT chain suggests the possibility of its presence in other types of classical many-body systems. Indeed, thermalization with an intermediate turbulent phase has also been recently observed in a system of one-dimensional hard rods \cite{DeNardis2024}.

\begin{figure*}[t]
\begin{subfigure}{0.46\textwidth}
		\includegraphics[width=\textwidth]{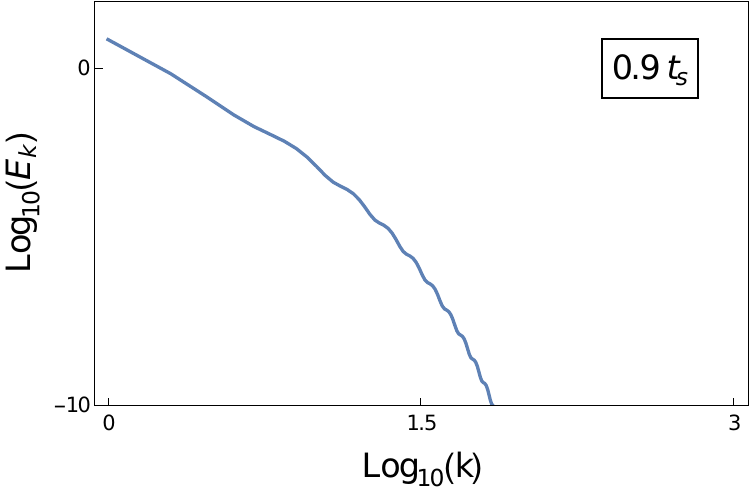} 
		\caption{}
\end{subfigure}
\begin{subfigure}{0.46\textwidth}
	\includegraphics[width=\textwidth]{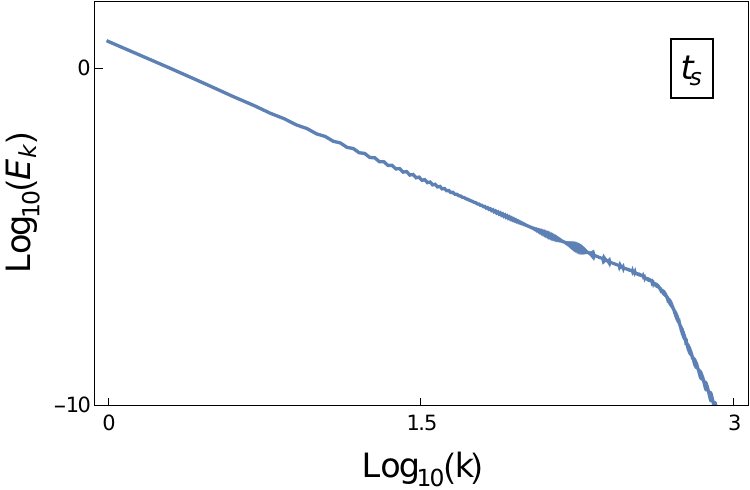}
	\caption{}
\end{subfigure}
	
\vspace{0.2cm}	
	
\begin{subfigure}{0.46\textwidth}
	\includegraphics[width=\textwidth]{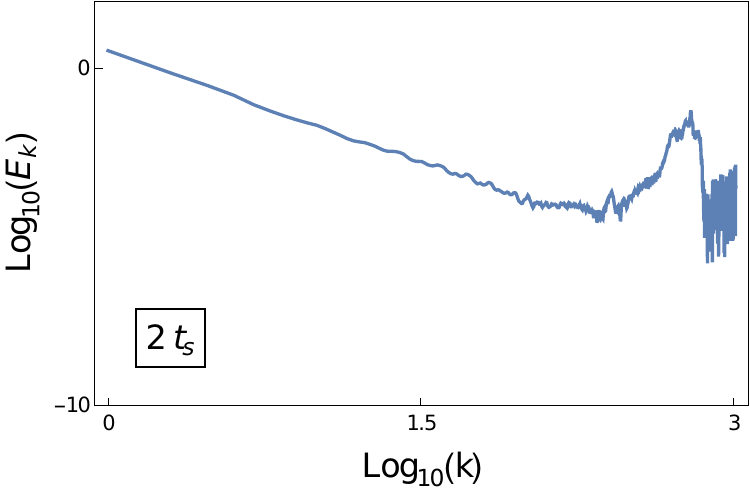}
	\caption{}
\end{subfigure}
\begin{subfigure}{0.46\textwidth}
	\includegraphics[width=\textwidth]{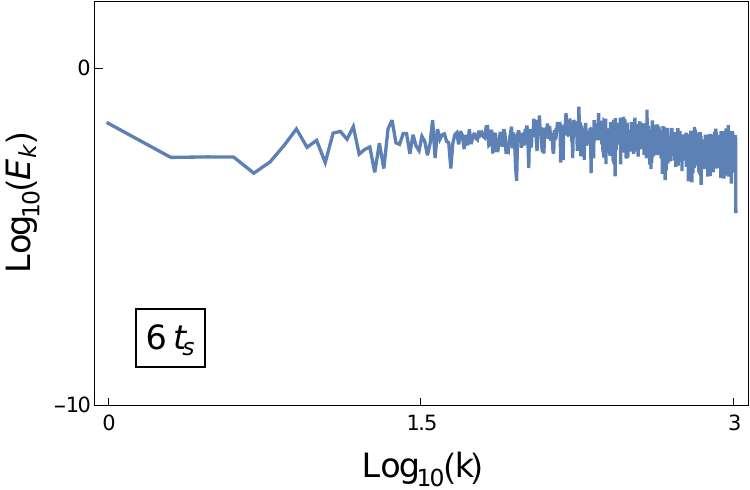}
	\caption{}
\end{subfigure}
				
		\caption{Fourier spectrum of the FPUT chain highlighting the power--law region. The panels show the spectrum at different times, measured in units of the shock time $t_s$ of the Burgers equation. Panel a: spectrum before the shock time. Panel b: spectrum at the shock time where the power--law slope is $\zeta=8/3$ as predicted by the Burgers dynamics. Panel c: spectrum in the turbulent range of the Burgers equation with $\zeta=2$. Panel d: spectrum at equipartition. The initial datum is a Traveling Wave Excitation \eqref{eq:indat} with $\varepsilon=0.05$, $\alpha=1$ and $\beta=0.5$.}\label{fig:Spectra}
\end{figure*}

\begin{figure}[t]
	\includegraphics[width=0.46\textwidth]{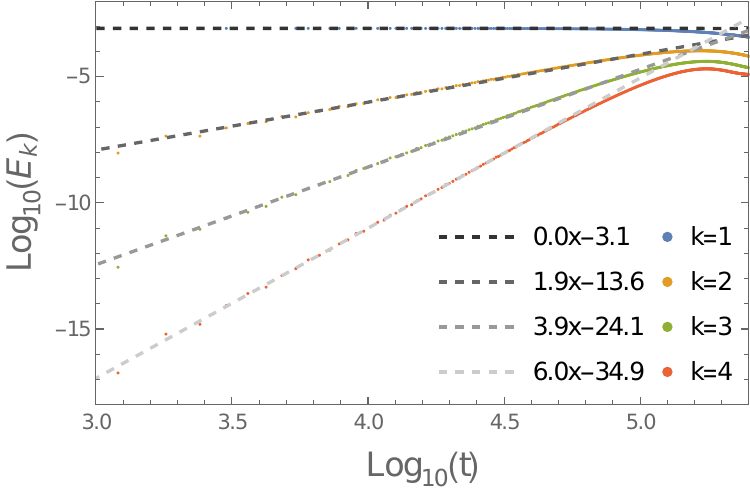}
	
	\caption{Initial time-evolution of the energies of the Fourier modes $k=1,2,3,4$. The energy of each Fourier mode grows with a power--law as predicted in \eqref{eq:InitialGrowthCC}. The numerical data are the points and the dashed lines represent the interpolation (see the legend in the inset). The initial condition is a traveling wave excitation with $\alpha=1$, $\beta=\frac12$, $\varepsilon=0.0005$, and $N=2048$.}
	\label{fig:Growth}
\end{figure}

In Sec.~\ref{sec:Results} we present the main results of the paper, we introduce the underlying continuum model and we discuss the thermodynamic limit. In Sec.~\ref{sec:PerturbationTheory} we introduce Hamiltonian perturbation theory and we use it to derive the pair of generalized Burgers equations directly from the FPUT lattice dynamics. In Sec.~\ref{sec:ShockTime} we derive the formula for the shock time for a class of initial conditions where a group of long--wavelength Fourier modes is excited. Finally, in Sec.~\ref{sec:Asymptotics} we derive the scaling for short times of the Fourier spectrum $E_k(t)$ and the large $k$ asymptotics of the Fourier spectrum at the shock time.

\section{Time--Evolution of the Spectrum and the Thermodynamic Limit}\label{sec:Results}
In this section we will present the time-evolution of a single mode initial condition and we will show how to perform the thermodynamic limit. We consider the  FPUT model \eqref{eq:H} with periodic boundary conditions: $q_N=q_0$ and $p_N=p_0$. 
Defining the Fourier coefficient $\hat{q}_k \;=\; \frac{1}{\sqrt{N}}\sum_{j=1}^N q_j e^{\imath 2 \pi k j/N}$ of the displacements $q_j$,
and similarly for the momenta $p_j$, the energy of the linearized system is consequently written as
\begin{equation}
		H_{lin}=\sum_{k=1}^N E_k \, ,  \qquad\, E_k:=\frac{|\hat{p}_k|^2}{2}+\frac{\omega_k^2 |\hat{q}_k|^2}{2} 
\end{equation}
where $\omega_k=2\sin(\pi k/N)$ and $E_k$ is the energy of mode $k$.

We consider the two-parameter family of initial data
\begin{subequations}
\label{eq:indat}
\begin{eqnarray}
	q_j(0)&=& A\cos\varphi\sin\left(\frac{2 \pi j}{N} \right)\ ;\\
	p_j(0)&=& \omega_1A \sin\varphi\cos\left(\frac{2 \pi j}{N}\right)\ ,	
\end{eqnarray}
\end{subequations}
for $j=1,\dots,N$. Here, $A>0$ and $0\leq\varphi\leq\pi/2$ are the amplitude and the phase of the initial excitation. Varying the phase from 
$\varphi=0$ to $\varphi=\pi/2$, we tune the kinetic energy of the initial condition (\ref{eq:indat}).
The value $\varphi=\pi/4$ corresponds to a left traveling wave excitation (TWE). This choice of the initial condition is motivated by the foundational studies of Zabusky and Kruskal of the Korteweg--de Vries (KdV) equation \cite{ZK65}. Around this initial condition, we explore a large neighborhood of states by varying both the amplitude and the phase. For this choice of initial condition, the specific energy $\varepsilon=E/N$ can be written in terms of $A$ and $\varphi$, for large $N$, as
\begin{equation}
\label{eq:epsa}
\varepsilon=a^2+\frac{3\beta}{2}\left(a\cos\varphi\right)^4\ ,\ \ a=\frac{\pi A}{N}\ .
\end{equation}
In presence of dispersion, this initial condition would give rise to a train of solitons \cite{ZK65} whose typical width is related to the strength of the dispersion. It is remarkable that the KdV equation describing the FPUT chain becomes dispersionless in the thermodynamic limit \cite{GP22}. Therefore, we expect a substantially different time evolution in this limit. Setting the model parameters in order to be close to this limit we show that the continuous field representing the lattice system in the limit develops a shock which is described by a (pair of) generalized Burgers equation. Both the time-evolution of the field and the shock time can be derived analytically. In Fig.~\ref{fig:Spectra} we represent the FES in double logarithmic scale just before the shock time (panel a), at the shock time (panel b), at twice the shock time (panel c) and at equipartition (panel d). In panel a the spectrum decays exponentially for large wave numbers as expected from previous studies \cite{Po03}. In panel b, the spectrum is well--fitted by
\begin{equation}\label{eq:fesgen2}
	\frac{E_k(t_s)}{E_{tot}} \sim 0.779 \, k^{-8/3} \, 
\end{equation}
for a large window of intermediate modes; this is exactly the spectrum at the shock time. In panel c we show that the power--law scaling persists after the shock time with an exponent $\zeta=2$ which persists for several shock times as we will show in the following. 

Another important prediction of the theory we develop in this paper concerns the energy cascade from long to shorter wavelength modes. One can derive analytically that the energy in the $k$-th mode, for short times, grows in time like
\begin{equation}
	E_k(t) \sim t^{2k-2} \, , \qquad t \ll t_s \, .
\end{equation}
In Fig. 3 we show the validity of this prediction for modes $k=1,\dots,4$. This is again a consequence of the underlying Burgers dynamics.

\subsection{Continuum model}\label{subsec:ContinuousModel}
The aim of this section is to embed the discrete FPUT system \eqref{eq:H} into a space-continuous model. Indeed, the discrete FPUT dynamics can be viewed as a sampling of an appropriate underlying continuous dynamics, the latter being generated by the equations of motion in the thermodynamic limit. With this idea in mind, we consider the variables $q_j(t)$ and $p_j(t)$ as a suitably rescaled sampling of the fields $Q(x,\tau)$ and $P(x,\tau)$ defined on a continuous spatial domain. In order to obtain a well-defined (and finite) specific energy in the limit, one has to choose
\begin{eqnarray}
	q_j(t)&=&NQ(j/N,t/N) \, , \\
	p_j(t)&=&P(j/N,t/N) \, .
\end{eqnarray}
We define $x_j=j/N$ and $\tau=t/N$, with this latter scaling set in order to obtain a finite time--scale for the propagation of a linear wave over the whole lattice. The derivation of the equations of motion for the fields $Q$ and $P$ is described in the following.
First, notice that 
\[
\begin{split}
q_{j+1}-q_j=&  Q_x(x_j,\tau)+Q_{xx}(x_j,\tau)/(2N)+\cdots \\
q_j-q_{j-1}=&  Q_x(x_j,\tau)-Q_{xx}(x_j,\tau)/(2N)+\cdots \, .\\
\end{split}
\]
Then, omitting the dependence on $\tau$, for any smooth function $f$, one has
\[
	\begin{split}
f(q_{j+1}-q_j)-f(q_j-q_{j-1})&=f'(Q_x(x_j))Q_{xx}(x_j)/N+\cdots\\
&=\frac{1}{N}[f(Q_x(x))]_x\Big|_{x=x_j}+\cdots\ ,
	\end{split}
\]
the dots denoting everywhere terms of higher order in $1/N$. 

Starting from the Hamiltonian \eqref{eq:H} we define the specific energy functional
\begin{equation}\label{eq:Limiting}
	\begin{split}
w[Q,P]&=\lim_{N\to\infty}\frac{H}{N}\\
&=\lim_{N\to\infty}\sum_{j=1}^N[P(x_j,\tau)+V(Q_x(x_j,\tau)+\cdots)]\Delta x_j\ ,
	\end{split}
\end{equation}
with $\Delta x_j=1/N$. Taking into account the expansions reported above, keeping $x_j$ finite and renaming it $x$, one finds the continuum form of the Hamiltonian in the limit $N\to\infty$
\begin{equation}
w[Q,P]=\oint\left[\frac{P^2}{2}+V(Q_x)\right]\ dx\ . \label{eq:w}
\end{equation}
The equations of motion associated to \eqref{eq:w} are
\begin{equation}
\begin{split}
Q_\tau&=\frac{\delta w}{\delta P}=P\, ,\\
 P_\tau&=-\frac{\delta w}{\delta Q}=[V'(Q_x)]_x\ .
\end{split} \label{eq:weqn}
\end{equation}
The integral $\oint$ in (\ref{eq:w}) is performed on the unit interval ($0<x\leq 1$) with periodic boundary conditions for the fields
$Q$ and $P$, and the functional derivatives $\delta w/\delta Q$ and $\delta w/\delta P$ are defined in the usual way \cite{Ref1}.

In the limit $N\to \infty$, the initial conditions (\ref{eq:indat}) read
\begin{equation}
\label{eq:inQP}
\begin{split}
Q(x,0)&=\frac{a}{\pi}\cos\varphi\sin(2\pi x)\ , \\
 P(x,0)&=2a\sin\varphi\cos(2\pi x)\ ,
\end{split}
\end{equation}
with $a \equiv \pi A/N$ supposed to be fixed and small in the limit $A\to\infty$, $N\to\infty$. By substituting the initial condition (\ref{eq:inQP})
into (\ref{eq:w}), and taking into account that $w$ is a constant of motion, i.e. it is preserved by the flow generated by Eqs.~(\ref{eq:weqn}), we obtain $w=\varepsilon$ for all times. As it can be explicitly checked, one gets 
\begin{equation}
	\varepsilon=a^2+\frac{3 \beta}{2} (a \cos \varphi)^4 \, , \quad a=\frac{\pi A}{N}
\end{equation}
which relates the amplitude $a$ to the specific energy $\varepsilon$ of the system. The latter quantity, and $a$ as a consequence, are supposed to be small. 

If we neglect the nonlinearity in $V$, Eqs. \eqref{eq:weqn} reduce to the wave equations for the field $Q$: 
\begin{equation}
	Q_{\tau \tau} = Q_{xx} \, .
\end{equation}
The dynamics of the linear model, in this continuum limit, is a superposition of left and right traveling waves on the unit circle. Since the nonlinearity is thought to be small, we study perturbatively the Hamiltonian initial value problem (\ref{eq:w})-- (\ref{eq:inQP}). We introduce the left ($L$) and right ($R$) rescaled fields 
\begin{equation}
\label{eq:LR}
\begin{split}
L(x,\tau)&=\frac{Q_x(x,\tau)+P(x,\tau)}{a\sqrt{2}}\ , \\
R(x,\tau)&=\frac{Q_x(x,\tau)-P(x,\tau)}{a\sqrt{2}}\ ,
\end{split}
\end{equation}
in terms of which the rescaled Hamiltonian functional $h[L,R]=\lim_{N\to\infty}H/(Na^2)$ takes the form 
\begin{equation}
\label{eq:h}
h=\frac{\langle L^2+R^2\rangle}{2}+\frac{a\alpha\langle(L+R)^3\rangle}{6\sqrt{2}}+
\frac{a^2\beta\langle(L+R)^4\rangle}{16}\ .
\end{equation}

Here and henceforth $\langle f\rangle\equiv\int_0^1 f(x)dx$ denotes the spatial average of $f(x)$. The Hamilton equations 
associated with (\ref{eq:h}) are obtained as $L_\tau=(\delta h/\delta L)_x$ and $R_\tau=-(\delta h/\delta R)_x$:
\begin{equation}
\label{eq:LRsys}
\left\{
\begin{split}
L_\tau=&\left[L+\frac{\alpha a}{2\sqrt{2}}(L+R)^2+
\frac{\beta a^2}{4}(L+R)^3\right]_x \\ 
R_\tau=&-\left[R+\frac{\alpha a}{2\sqrt{2}}(L+R)^2+
\frac{\beta a^2}{4}(L+R)^3\right]_x
\end{split}\right.\ ,
\end{equation}
which can be obtained by direct substitution of (\ref{eq:LR}) into (\ref{eq:weqn}). In the same way, the initial conditions (\ref{eq:inQP}) transform into
\begin{equation}
\label{eq:indatLR}
\left\{
\begin{split}
L_0(x)\equiv L(x,0)=&\sqrt{2}(\cos\varphi+\sin\varphi)\cos(2\pi x) \\
R_0(x)\equiv R(x,0)=&\sqrt{2}(\cos\varphi-\sin\varphi)\cos(2\pi x)
 \end{split}
 \right.\ .
\end{equation}
Notice that $L_0$ has maximal amplitude $2$ for $\varphi=\pi/4$, when $R_0=0$, which is the left TWE. We also observe that $\oint P\ dx=\langle P\rangle$ is a constant of motion of the system (\ref{eq:weqn}). From the initial condition (\ref{eq:inQP}) it follows that $\langle P\rangle=0$, which in turn implies the conservation law $\langle Q \rangle=0$. Then, by integrating (\ref{eq:LR}), one gets $\langle L\rangle=0$ and $\langle R\rangle=0$.

\subsection{Relation with the thermodynamic limit}

In order to define the dynamics in the thermodynamic limit one needs to perform the limits $E \to +\infty$, $N\to +\infty$ in the equations of motion by keeping the ratio $E/N=\varepsilon$ fixed. This amounts to performing the limit $N\to+\infty$ in \eqref{eq:Limiting} by keeping $\varepsilon$ in \eqref{eq:epsa} fixed. 

In the limiting procedure, initial data \eqref{eq:inQP} remain unchanged, as $a$ is an intensive parameter depending only on $\varepsilon$, $\varphi$, $\alpha$ and $\beta$. The limiting procedure for the equations is more complicated: neglecting higher order derivatives of the field $Q$ in $q_{j+1}-q_j$, is a delicate point. Indeed, keeping all the term as is done in \cite{GP22} would yield a pair of counter-propagating KdV equations instead of \eqref{eq:LRsys} with a coefficient proportional to $N^{-2}$ in front of the term with three derivatives. Then, the thermodynamic limit is encoded in a zero-dispersion limit for that (pair of) KdV equations. Before the shock time, the procedure just described yields the pair of generalized Burgers equations \eqref{eq:LRsys}. The problem is still open for times larger than the shock time. See Sec.~\ref{subsec:AfterTS} for a more detailed discussion on the zero-dispersion limit.

\section{Perturbation theory: derivation of the decoupled generalized Burgers equations}\label{sec:PerturbationTheory}

\subsection{Review of general theory}

Let us fix a generic Hamiltonian functional $w[Q,P]=\oint \mathcal{W} dx=\langle \mathcal{W}\rangle$ whose density $\mathcal{W}$ is a function of the fields 
$Q_x$, $P$ and of their derivatives with respect to $x$ up to a given finite order. In the following, functionals are denoted by lowercase letters, whereas their densities are denoted 
by the same capital, calligraphic letter. The FPUT Hamiltonian (\ref{eq:w}) belongs to this class of functionals. Let us then consider a functional 
$f[Q,P]=\oint \mathcal{F} dx=\langle \mathcal{F}\rangle$ whose density $\mathcal{F}$ is in the same class of $\mathcal{W}$. Then, given the Hamilton equations
\begin{equation}
\label{eq:caneq}
Q_\tau=\frac{\delta w}{\delta P}\ \ ;\ \ P_\tau=-\frac{\delta w}{\delta Q}\ ,
\end{equation}
the time derivative of $f$ along their solution satisfies
\begin{equation}
\label{eq:dfdt}
\frac{df}{d\tau}=\left\langle \frac{\delta f}{\delta Q}\frac{\delta w}{\delta P}-\frac{\delta f}{\delta P}\frac{\delta w}{\delta Q}\right\rangle\equiv
\{f,w\}_c\ ,
\end{equation}
where we have defined the canonical Poisson bracket $\{\ ,\}_c$ relative to the canonical coordinates $Q$ and $P$. One can check directly that $\{ \, , \, \}_c$  is bilinear, antisymmetric and satisfies Jacobi identity and Leibnitz rule and thus it is indeed a Poisson bracket. The equations of motion (\ref{eq:caneq}) read $Q_\tau=\{Q,w\}_c$, $P_\tau=\{P,w\}_c$. Everything in this procedure is completely analogous to the finite dimensional case. 

By applying the transformation (\ref{eq:LR}), one has $w[Q,P]\to h[L,R]=\langle \mathcal{H}\rangle$, 
$f[Q,P]\to \tilde f[L,R]=\langle\tilde{\mathcal{F}}\rangle$, where their respective densities 
$\mathcal{H}$ and $\tilde{\mathcal{F}}$ depend on $L$, $R$ and their derivatives with respect to $x$ up to a finite order. As a consequence, formula (\ref{eq:dfdt}) transforms into
\begin{equation}
\label{eq:dftildt}
\frac{d\tilde f}{d\tau}=
\left\langle \frac{\delta \tilde f}{\delta L}\left(\frac{\delta h}{\delta R}\right)_x-
\frac{\delta \tilde f}{\delta R}\left(\frac{\delta w}{\delta L}\right)_x\right\rangle\equiv
\{\tilde f,h\}_G\ ,
\end{equation}
where this last equation defines the Gardner bracket $\{\ ,\}_G$. Since (\ref{eq:dfdt}) and (\ref{eq:dftildt}) express the time derivative of one and the same functional $f$ (or $\tilde f$), they imply the identity $\widetilde{\{f,w\}}_c=\{\tilde f,h\}_G$. 
The tilde on the left means that one first computes the Poisson bracket of $f$ and $w$ with respect to $Q$ and $P$ and then performs the change of variables (\ref{eq:LR}) from $(Q,P)$ to $(L,R)$. Thus, the bracket defined in (\ref{eq:dftildt}) is also a Poisson bracket, being just the transformed bracket of the canonical Poisson bracket defined in (\ref{eq:dfdt}). The equations of motion
associated to $h$ in the latter structure read $L_\tau=\{L,h\}_G$, $R_\tau=\{R,h\}_G$, that for the FPUT Hamiltonian (\ref{eq:h}) yields just the equations (\ref{eq:LRsys}).

We define the operator $\mathcal{L}_h=\{\ ,h\}_G$ which acts on $\tilde f$ as
$\mathcal{L}_h\tilde f=\{\tilde f,h\}_G$. Equation (\ref{eq:dftildt}) can be symbolically solved by 
\begin{equation}
\label{eq:ftiltau}
\begin{split}
\tilde f(\tau)&=e^{\mathcal{L}_h\tau}\tilde f(0)\\&=\left(1+\tau\mathcal{L}_h+\frac{\tau^2}{2}\mathcal{L}_h^2+\cdots\right)
\tilde f(0)\ ,
\end{split}
\end{equation}
where the exponential of $\tau\mathcal{L}_h$ is defined by its series, the dots denoting higher order terms. The fundamental result used below is the following: the flow $e^{\tau\mathcal{L}_h}$ of $h$, with $\mathcal{L}_h=
\{\ ,h\}_G$, preserves the Poisson bracket $\{\ ,\}_G$ itself, in the sense that
$e^{\tau\mathcal{L}_h}\{a,b\}_G=\left\{e^{\tau\mathcal{L}_h}a,e^{\tau\mathcal{L}_h}b\right\}_G$
for any pair of functionals $a=\langle \mathcal{A}\rangle$ and $b=\langle \mathcal{B}\rangle$ in the given class
\cite{Ref2}.
As a final remark, notice that in the above treatment no special form of $h$ (and the other functionals) was considered, so that the flow of \emph{any} functional $h$ preserves the Hamiltonian structure defining it. This derivation is completely general, and the Gardner structure is just an example. 

\subsection{Perturbative analysis of the continuum FPUT model}
We now make use of the above tools to transform the Hamiltonian (\ref{eq:h}) of the continuum FPUT model and decouple the equation for the left $L$ and right $R$ fields in the corresponding Hamilton equations to second order in the small parameter $a$. The Hamiltonian (\ref{eq:h}) can be obviously ordered as
\begin{equation}\label{eq:h0+1+2}
h=h_0+ah_1+a^2h_2
\end{equation}
where 
\begin{equation}
\begin{split}
\label{eq:h012}
h_0&=\frac{\langle L^2+R^2\rangle}{2}\, , \\
h_1&=\frac{\alpha\langle(L+R)^3\rangle}{6\sqrt{2}}\, , \\ 
h_2&=\frac{\beta\langle(L+R)^4\rangle}{16}\ .
\end{split}
\end{equation}
The equations of motion of $h_0$ have the form $L_\tau=L_x$, $R_\tau=-R_x$ (see \eqref{eq:LRsys}), whose solution is
$L(x,\tau)=L_0(x+\tau)$, $R(x,\tau)=R_0(x-\tau)$, i.e.\ the left and right translation of the initial condition, respectively.
The latter solution is periodic in time, with spatial period one, for any space periodic initial condition $(L_0,R_0)$ with period one. Then, with the notation introduced above, the flow of $h_0$ at time $s$ is defined by
\begin{equation}
\label{eq:flow0}
(L_0(x+s),R_0(x-s))=e^{s\mathcal{L}_0}(L_0(x),R_0(x) \, ,
\end{equation}
with $\mathcal{L}_0$  given by
\begin{equation} 
\mathcal{L}_0=\{\ ,h_0\}_G\ .
\end{equation}
Since the flow of $h_0$ has period one, $e^{\mathcal{L}_0}=1$. 

We now build up a transformation of the fields $\mathcal{C}_a:(L,R)\mapsto(\lambda,\rho)=\mathcal{C}_a(L,R)$, smoothly dependent on the small parameter $a$ and close to the identity ($\mathcal{C}_0(L,R)=(L,R)$), by composing two Hamiltonian flows corresponding to two unknown generating Hamiltonians, $g_1$ and $g_2$, as follows.  
By defining $\mathcal{L}_1=\{\ ,g_1\}_G$, $\mathcal{L}_2=\{\ ,g_2\}_G$, we set
\begin{equation}
\label{eq:Cacomp}
(L,R)=\mathcal{C}^{-1}_a(\lambda,\rho)=e^{a^2\mathcal{L}_2}e^{a\mathcal{L}_1}(\lambda,\rho)\ .
\end{equation}
The conditions below uniquely determine $g_1$ and $g_2$ \cite{GP22}.
\begin{enumerate}
\item The transformed Hamiltonian $\tilde h=h\circ \mathcal{C}_a^{-1}=e^{a^2\mathcal{L}_2}e^{a\mathcal{L}_1}h$
is in normal form with respect to $h_0$ to second order in $a$, namely
$\tilde h=h_0+a \tilde{h}_1 + a^2\tilde{h}_2+O(a^3)$, with $\{\tilde{h}_1,h_0\}_G=\{\tilde{h}_2,h_0\}_G=0$ (i.e.
$\tilde{h}_1$ and $\tilde{h}_2$ are first integrals of $h_0$).
\item $g_1$ and $g_2$ have zero average on the unperturbed flow of $h_0$: 
$\int_0^1e^{s\mathcal{L}_0}g_1 ds=\int_0^1e^{s\mathcal{L}_0}g_2 ds=0$.
\end{enumerate}
In order to write explicitly the Taylor series with respect to the small parameter $a$, we expand the exponentials $e^{a^2 \mathcal{L}_2} e^{a \mathcal{L}_1}$ and apply them to $h$ given by \eqref{eq:h0+1+2}. We get
\begin{eqnarray}
\tilde h&=&e^{a^2\mathcal{L}_2}e^{a\mathcal{L}_1}(h_0+a h_1+a^2 h_2)=\nonumber\\
&=& h_0+a(\mathcal{L}_1h_0+h_1)+ \\
& & \quad+
a^2\left(\mathcal{L}_2h_0+\mathcal{L}_1h_1+\frac{1}{2}\mathcal{L}_1^2h_0+h_2\right)+O(a^3)=\nonumber\\
&=&h_0+a \tilde{h}_1 + a^2\tilde{h}_2+O(a^3)\ .\nonumber
\end{eqnarray}
Taking into account that $\mathcal{L}_1h_0=-\mathcal{L}_0g_1$ and $\mathcal{L}_2h_0=
-\mathcal{L}_0g_2$, one finds the two \emph{homological equations} (see  \cite{Giorgilli} and references therein)
\begin{equation}
\label{eq:homo12}
\begin{split}
\mathcal{L}_0g_1= & h_1-\tilde{h}_1\ ; \\
\mathcal{L}_0g_2= & \mathcal{L}_1h_1+\frac{1}{2}\mathcal{L}_1^2h_0+h_2-\tilde{h}_2\ ,
\end{split}
\end{equation}
for the four unknowns $g_1$, $g_2$, $\tilde{h}_1$ and $\tilde{h}_2$. The solution of Eqs.~(\ref{eq:homo12}) can be obtained taking into account the following technical points. First, since $h_0$ has a periodic flow with period one, $e^{\mathcal{L}_0}=1$ and, therefore,
\begin{equation}
\label{eq:step1}
\begin{split}
\int_0^1e^{s\mathcal{L}_0}\mathcal{L}_0g_i\ ds&=\int_0^1\frac{d}{ds}e^{s\mathcal{L}_0}g_i\ ds\\
&=\left(e^{\mathcal{L}_0}-1\right)g_i=0 
\end{split}
\end{equation}
for $i=1,2$. Second, since $\tilde{h}_i$ is in normal form with respect to $h_0$, we have $\mathcal{L}_0\tilde{h}_i=\{\tilde{h}_i,h_0\}_G=0$, and therefore
\begin{equation}
\label{eq:step2}
e^{s\mathcal{L}_0}\tilde{h}_i=\left(1+s\mathcal{L}_0+\frac{s^2}{2}\mathcal{L}_0^2+\cdots\right)\tilde{h}_i= 
\tilde{h}_i
\end{equation}
for $i=1,2$. Third, for $i=1,2$,
\begin{equation}
\label{eq:step3}
\int_0^1se^{s\mathcal{L}_0}\mathcal{L}_0g_i\ ds=se^{s\mathcal{L}_0}g_i\Big|_{0}^1
-\int_0^1e^{s\mathcal{L}_0}g_i\ ds=g_i\ ,
\end{equation}
 since we have imposed that the average of the generating Hamiltonians $g_i$ on the unperturbed flow vanishes. 
 
By taking into account the steps (\ref{eq:step1}), (\ref{eq:step2}) and (\ref{eq:step3}), one obtains the solution of the first of Eqs.~(\ref{eq:homo12}), namely
\begin{equation}
\label{eq:htil1g1}
\tilde{h}_1=\int_0^1e^{s\mathcal{L}_0}h_1\ ds\ \ ;\ \ 
g_1=\int_0^1 se^{s\mathcal{L}_0}(h_1-\tilde{h}_1)\ ds\ .
\end{equation}

Before solving the second of the Eqs.~(\ref{eq:homo12}) in an analogous way, it is convenient to substitute 
$\mathcal{L}_1h_0=\tilde{h}_1-h_1$ into its right hand side, to get
$\mathcal{L}_0g_2= \frac{1}{2} \mathcal{L}_1h_1+h_2-\tilde{h}_2+\frac{1}{2}\mathcal{L}_1\tilde{h}_1$. Now,
the average of $\mathcal{L}_1\tilde{h}_1$ vanishes:
\[
\begin{split}
\int_0^1 e^{s\mathcal{L}_0}\mathcal{L}_1\tilde{h}_1\ ds&=\int_0^1 e^{s\mathcal{L}_0}\{\tilde{h}_1,g_1\}_G\ ds=\\
&=
\left\{\tilde{h}_1,\int_0^1e^{s\mathcal{L}_0}g_1ds\right\}_G=0\ ,
\end{split}
\]
by (\ref{eq:step2}) and the bilinearity of the Poisson bracket. Thus, the average of the second of Eqs.~(\ref{eq:homo12}) yields
\begin{equation}
\label{eq:htil2}
\begin{split}
\tilde{h}_2%&=\int_0^1e^{s\mathcal{L}_0}\left(h_2+\frac{1}{2}\mathcal{L}_1h_1\right)ds\\
&=\int_0^1e^{s\mathcal{L}_0}\left(h_2+\frac{1}{2}\{h_1,g_1\}_G\right)ds\ .
\end{split}
\end{equation}

We do not report here the expression of the generating Hamiltonian $g_2$, since it is not used for the computation of the Hamiltonian to second order, nor is it useful for the transformation of initial data.

One has now to explicitly compute the quantities (\ref{eq:htil1g1}) and (\ref{eq:htil2}), where the 
functions $h_1$ and $h_2$ given in (\ref{eq:h012}), have to be expressed in the new field variables $\lambda(x,\tau)$ and $\rho(x,\tau)$. By its definition (\ref{eq:flow0}), the action of 
$e^{s\mathcal{L}_0}$ on a monomial in $\lambda$
and $\rho$ is:
\[
e^{s\mathcal{L}_0}\lambda^m(x,\tau)\rho^n(x,\tau)=\lambda^m(x+s,\tau)\rho^n(x-s,\tau)\ .
\] 
In order to explicitly compute $\tilde{h}_1$, $g_1$ and $\tilde{h}_2$ one needs the following relations, valid for any pair of functions $F$ and $G$ of spatial period one:
\[
\int_0^1\int_0^1F(x+s)G(x-s)\ dx\ ds=\int_0^1F(x)\ dx\ \int_0^1 G(x)\ dx\ ;
\]
\begin{equation}
\begin{split}
&\int_0^1\int_0^1s\ F(x+s)G(x-s)\ dx\ ds=\\
&\frac{1}{2}\left(\int_0^1F(x)\ dx\ \int_0^1 G(x)\ dx+
\int_0^1G(x)\partial_x^{-1} F(x)\ dx\right)\ ,
\end{split}
\end{equation}
which can be proven by expressing $F$ and $G$ in Fourier series. The antiderivative $\partial_x^{-1}$ appearing above is defined as $\partial_x^{-1}F(x)=\sum_{k\neq0}\hat F_k/(\imath2\pi k)e^{\imath2\pi kx}$. Observe that the antiderivative is skew-symmetric under integration.

After elementary, though a bit long computations (that involve also the explicit determination of $g_1$; see (\ref{eq:g1}) below), one finds the explicit expression of the normal form Hamiltonian 
$\tilde{h}=h_0+a\tilde{h}_1 +a^2 \tilde{h}_2+O(a^3)$, namely
\begin{equation}
\label{eq:htilrep}
\begin{split}
\tilde h=&\frac{\langle\lambda^2+\rho^2\rangle}{2}+a\left(\alpha \frac{\langle\lambda^3+
\rho^3\rangle}{6\sqrt{2}}\right) + \\
& +a^2\Bigg[\left(2\beta-\alpha^2\right)\frac{\langle\lambda^4+\rho^4\rangle}{32}
+\alpha^2\frac{\langle\lambda^2\rangle^2+\langle\rho^2\rangle^2}{32} \\
&+\left(3\beta-2\alpha^2\right)\frac{\langle\lambda^2\rangle\langle\rho^2\rangle}{8}\Bigg]+O(a^3)\ .
\end{split}
\end{equation}
The equations of motion associated with this Hamiltonian, up to terms $O(a^3)$, are
\begin{equation}
\label{eq:larhoev}
\left\{
\begin{split}
& \lambda_\tau=
\left[c_l+\frac{a\alpha}{\sqrt{2}}\lambda+
\frac{3a^2\alpha^2}{4}\left(\frac{\beta}{\alpha^2}-\frac{1}{2}\right)\lambda^2\right]\lambda_x \\
& \rho_\tau=
-\left[c_r+\frac{a\alpha}{\sqrt{2}}\rho+
\frac{3a^2\alpha^2}{4}\left(\frac{\beta}{\alpha^2}-\frac{1}{2}\right)\rho^2\right]\rho_x
\end{split}
\right.\ ,
\end{equation} 
where the left and right translation velocities $c_l$ and $c_r$ are given by
\begin{equation}
\begin{split}
& c_l=1+\frac{\alpha^2a^2}{8}\langle\lambda^2\rangle+
\frac{3a^2}{4}\left(\beta-\frac{2\alpha^2}{3}\right)\langle\rho^2 \rangle\ ; \\
& c_r=1+\frac{\alpha^2a^2}{8}\langle\rho^2\rangle+
\frac{3a^2}{4}\left(\beta-\frac{2\alpha^2}{3}\right)\langle\lambda^2 \rangle\ .
\end{split}
\end{equation}
Notice that $\langle \lambda^2\rangle$ and $\langle \rho^2\rangle$ are first integrals of system
(\ref{eq:larhoev}), so that the above velocities are constant. Moreover, the field transformation defined by
$\lambda(x,\tau)=\lambda'(x+c_l\tau,\tau)$, $\rho(x,\tau)=\rho'(x-c_r\tau,\tau)$, removes the translation terms 
$c_l\lambda_x$ and $-c_r\rho_x$ in system (\ref{eq:larhoev}), which completely decouples the two equations.
We thus set $c_l=c_r=0$ without any loss of generality (which also amounts to erase all the terms containing 
$\langle\lambda^2\rangle$, $\langle\rho^2\rangle$ and their powers in the Hamiltonian \eqref{eq:htilrep}).  
It is worth remarking that this transformation removes the coupling up to order $O(a^3)$.

Concerning the initial conditions $(\lambda_0,\rho_0)$ satisfied by the fields $\lambda$ and $\rho$, one deduces them as follows.
The first generating Hamiltonian $g_1$, defining the canonical transformation to first order in $a$, and necessary
to compute $\tilde{h}_2$, is 
\begin{equation}
\label{eq:g1}
g_1=\frac{\alpha}{4\sqrt{2}}\ \left\langle\rho\partial_x^{-1}(\lambda^2)+\rho^2\partial_x^{-1}\lambda\right\rangle\ .
\end{equation} 
One can then express the new fields $\lambda$ and $\rho$ in terms of the old ones, $L$ and $R$, by inverting 
(\ref{eq:Cacomp}), namely 
\[
\begin{split}
\lambda&= e^{-a\mathcal{L}_1}L=L-a\{L,g_1\}+O(a^2)\ , \\ 
\rho&= e^{-a\mathcal{L}_1}R=R-a\{R,g_1\}+O(a^2)\ .
\end{split}
\]
Neglecting a remainder of $O(a^2)$, the final result is
\begin{equation}
\label{eq:Caoa}
\begin{split}
\lambda=&  L+\frac{\alpha a}{4\sqrt{2}}\left(R^2-\langle R^2\rangle\right)+
\frac{\alpha a}{2\sqrt{2}}(LR+L_x\partial_x^{-1}R)\ ;\\
\rho=& R+\frac{\alpha a}{4\sqrt{2}}\left(L^2-\langle L^2\rangle\right)+
\frac{\alpha a}{2\sqrt{2}}(LR+R_x\partial_x^{-1}L)\ .
\end{split}
\end{equation} 
Substituting in the latter expression the initial condition (\ref{eq:indatLR}), with $\theta=\varphi-\pi/4$, and neglecting the remainder $O(a^2)$, one gets
\begin{equation}
\label{eq:indatlarho}
\left\{
\begin{split}
\lambda_0= & 2\cos\theta\cos(2\pi x)+\frac{a\alpha(\sin^2\theta-2\sin2\theta)}{2\sqrt{2}}\cos(4\pi x)\\
\rho_0 = &-2\sin\theta\cos(2\pi x)+\frac{a\alpha(\cos^2\theta-2\sin2\theta)}{2\sqrt{2}}\cos(4\pi x)
\end{split}
\right.\ .
\end{equation}
We finally observe that the transformation (\ref{eq:Caoa}) preserves the space average of the fields, so that $\langle \lambda\rangle=\langle L\rangle=0$ and $\langle \rho\rangle=\langle R\rangle=0$.

\section{Computation of the Shock Time}\label{sec:ShockTime}

Our transformed system (\ref{eq:larhoev}), (\ref{eq:indatlarho}), has the form of two decoupled, generalized inviscid Burgers equations. From now on, we will omit the term inviscid for simplicity. The solution of the generalized Burgers equation $u_\tau=f(u)u_x$, with initial datum $u_0(x)$, is implicitly defined by the equation $u-u_0(x+f(u)\tau)=0$, which can be checked by direct inspection. The latter identity admits an explicit solution if the implicit function theorem applies, namely, taking the derivative with respect to $u$, if
\[
1-u_0'(x+f(u)\tau)f'(u)\tau=1-\tau\frac{d}{d\xi}f(u_0(\xi))\neq0\ ; 
\]
where $\xi\equiv x+f(u)\tau$.
The above condition is satisfied for all $\tau$'s in the interval $[0,\tau_s[$, where $\tau_s$, the shock time, is given by 
\begin{equation}
\label{eq:16}
\frac{1}{\tau_s}= \max_x\frac{d}{dx}f(u_0(x))\ .
\end{equation}
The system \eqref{eq:larhoev} consists of two independent equations of the form $\lambda_\tau=f(\lambda) \lambda_x$ and $\rho_\tau=f(\rho)\rho_x$. We define the shock time of the FPUT system as $\tau_s=\min\{\tau_s^l,\tau_s^r\}$, whereas the left and right shock times $\tau^l_s$ and $\tau_s^r$ are given by
\begin{equation}
\label{eq:taulr}
\begin{split}
\frac{1}{\tau_s^l}&=\max_{x\in[0,1]}\left[\frac{d}{dx}\Phi(\lambda_0(x))\right]\ \ ;\\ 
\frac{1}{\tau_r^l}&=\max_{x\in[0,1]}\left[-\frac{d}{dx}\Phi(\rho_0(x))\right]\ .
\end{split}
\end{equation}
Here $\Phi$ is the function defined as
\begin{equation}\label{eq:Phi}
	\Phi(\lambda):=\frac{a \alpha}{\sqrt{2}} \lambda+\frac{3 a^2 \alpha^2}{4} \left(\frac{\beta}{\alpha^2}-\frac{1}{2} \right) \, ,
\end{equation}
where $\lambda_0$ and $\rho_0$ are given in (\ref{eq:indatlarho})
and  in $\Phi(\lambda_0(x))$ and $\Phi(\rho_0(x))$ we have consistently neglected terms of order $O(a^3)$. The explicit computation of $\tau_s^l$ and $\tau_s^r$ in \eqref{eq:taulr} for the left shock time yields
\begin{equation}
\label{eq:taul}
\begin{split}
\tau_s^l&=\left(\frac{1}{2\pi\sqrt{2}a\alpha}\right)\frac{1}{\cos\theta} \times \\
&\times
\sqrt{\frac{32\mu^2}{\sqrt{1+32\mu^2}-1+16\mu^2}}\frac{4}{\sqrt{1+32\mu^2}+3}\ ,
\end{split}
\end{equation}
where 
\[
\mu=\frac{a\alpha}{2\sqrt{2}}\cos\theta\left[
\tan^2\theta-4\tan\theta+6\left(\frac{\beta}{\alpha^2}-\frac{1}{2}\right)\right]\ , 
\]
whereas the right shock time is given by
\begin{equation}
\label{eq:taur}
\begin{split}
\tau_s^r&=\left(\frac{1}{2\pi\sqrt{2}a\alpha}\right)\frac{1}{|\sin\theta|} \times \\
&\times
\sqrt{\frac{32\eta^2}{\sqrt{1+32\eta^2}-1+16\eta^2}}\frac{4}{\sqrt{1+32\eta^2}+3}\ ,
\end{split}
\end{equation}
where
\[
\eta=\frac{a\alpha}{2\sqrt{2}}\sin\theta\left[
\cot^2\theta-4\cot\theta+6\left(\frac{\beta}{\alpha^2}-\frac{1}{2}\right)\right]\ . 
\]
In the range $-\pi/4\leq\theta\leq\pi/4$, for $a$ small enough, and any $\alpha$, $\beta$, 
the inequality $\tau_s^l\leq\tau_s^r$ holds,
the equality being valid only for $\theta=\pm\pi/4$.   Recalling that $\tau=t/N$, it follows that in the same range of $\theta$ and $a$ and for any
$\alpha$, $\beta$, $\tau_s=\min\{\tau_s^l,\tau_s^r\}=\tau_s^l$, one gets 
\begin{equation}\label{eq:ts}
	t_s=\left( \frac{N}{2 \pi \sqrt{2} a \alpha} \right) \frac{F(\mu)}{\cos \theta} \, ,
\end{equation}
where
\begin{equation}\label{eq:Fmu}
	F(\mu)=\sqrt{\frac{32 \mu^2}{\sqrt{1+32\mu^2}-1+16 \mu^2}} \frac{4}{\sqrt{1+32\mu^2}+3}
\end{equation}
and
\begin{equation}\label{eq:mu}
	\mu=\frac{a \alpha}{2 \sqrt{2}} \cos \theta \left[\tan^2 \theta-4 \tan \theta+6\left(\frac{\beta}{\alpha^2}-\frac{1}{2} \right) \right] \, .
\end{equation}

\section{FES asymptotics}\label{sec:Asymptotics}

\subsection{Shock time}\label{subsec:ShockTime}
Given the generalized Burgers equation $u_\tau=f(u)u_x$, we first write its solution in Fourier series, namely $u(x,\tau)=\sum_k\hat{u}_k(\tau)e^{\imath 2\pi kx}$, where the Fourier coefficient $\hat u_k(\tau)$
can be expressed in terms of the initial condition $u_0(x)$ by the explicit formula
\begin{equation}\label{eq:uk}
	\hat{u}_k(\tau)=\frac{1}{2 \pi \imath k} \oint u'_0(x)e^{- \imath 2 \pi k[x-\tau \Phi(u_0(x))]} dx \, .
\end{equation}
Taking into account that $u=u_0(x+\tau f(u))$, and introducing the variable $\xi$ such that $\xi=x+\tau f(u_0(\xi))$, one has
\begin{equation}\label{eq:Cinquanta}
\begin{split}
\hat{u}_k(\tau)=& \oint e^{-\imath2\pi kx}u_0(x+f(u)\tau)\ dx \\
=&   \oint e^{-\imath2\pi k[\xi-\tau f(u_0(\xi))]} u_0(\xi) \frac{d}{d\xi}\left[\xi-\tau f(u_0(\xi))\right]\ d\xi \\
=& \frac{1}{-\imath2\pi k}\oint u_0(\xi)\frac{d}{d\xi}\left[e^{-\imath2\pi k[\xi-\tau f(u_0(\xi))]}\right]\ d\xi\\
=&
 \frac{1}{\imath2\pi k}\oint u'_0(\xi)e^{-\imath2\pi k[\xi-\tau f(u_0(\xi))]}\ d\xi\ .
\end{split}
\end{equation}
Using \eqref{eq:Cinquanta} in the first equation of
\eqref{eq:larhoev}, that is with $f=\Phi$, yields \eqref{eq:uk}.

Let us now assume that the initial datum $u_0(x)$ of the generalized Burgers equation
$u_\tau=f(u)u_x$ satisfies the following three conditions: 
\begin{enumerate}
\item[(i)] $u_0(x)=\sum_{n=-M}^Mc_ne^{\imath2\pi nx}$; $c_0=0$, $M$ finite;
\item[(ii)] $df(u_0(x))/dx$ admits a finite number $m$ of absolute maximum points $x_1,\dots,x_m\in[0,1[$;
\item[(iii)] $\gamma_j\equiv d^3f(u_0(x_j))/dx^3\neq0$.
\end{enumerate}

This is a generalization of the case where the first Fourier mode is excited. Indeed, in that case, the number of critical points is two, a maximum and a minimum.

We start  by the expression (\ref{eq:uk}) for $\hat u_k(\tau)$, and split the unit integration interval into $m$ disjoint subintervals $I_1,\dots,I_m$, such that $I_j$ contains only the maximum point $x_j$ in its interior. Thus
\begin{equation}
\label{eq:uktau}
u_k(\tau)=\frac{1}{k}\sum_{n=-M}^Mc_nn\sum_{j=1}^m\int_{I_j}e^{\imath2\pi nx}
e^{-\imath2\pi k[x-\tau f(u_0(x))]}\ dx\ .
\end{equation}
In the asymptotics $k\to\infty$ each of the integrals on $I_j$ is treated with the method of stationary phase \cite{Ref3}. One has to take into account that, by the definition (\ref{eq:16}) of the shock time $\tau_s$, and by the hypotheses (ii) and (iii) above, in the interval $I_j$
\[
	\begin{split}
x-\tau_s f(u_0(x))&=x_j-\tau_s f(u_0(x_j))-\frac{\tau_s\gamma_j}{6}(x-x_j)^3\\&+O((x-x_j)^4)\ .
	\end{split}
\]
Thus, if $I_j=[x_j-a_j,x_j+b_j[$, by changing variable to $u=x-x_j$, for $k\gg 1$ and $\tau=\tau_s$ one finds
\begin{equation}
\label{eq:Ij}
\begin{split}
&\int_{I_j}e^{\imath2\pi nx}e^{-\imath2\pi k[x-\tau_s f(u_0(x))]}\ dx \\%&= % e^{\imath2\pi nx_j}
%e^{-\imath2\pi k[x_j-\tau_s f(u_0(x_j))]}\int_{-a_j}^{b_j}e^{\imath2\pi n u}
%e^{\imath\frac{\pi \tau_s\gamma_j}{3}ku^3+O(ku^4)}\ du=\\
%&=\  \frac{e^{\imath2\pi nx_j}e^{-\imath2\pi k[x_j-\tau_s f(u_0(x_j))]}}{(\pi|\gamma_j|\tau_s k)^{1/3}}
%\int_{-a_j(\pi|\gamma_j|\tau_sk)^{1/3}}^{b_j(\pi|\gamma_j|\tau_s k)^{1/3}}
%e^{\imath \frac{2\pi n}{(\pi|\gamma_j|\tau_s k)^{1/3}}z}
%e^{\imath \mathrm{sgn}(\gamma_j)\frac{z^3}{3}+O(z^4/k^{1/3})}\ dz\sim \\
%\sim \  \frac{e^{\imath2\pi nx_j}e^{-\imath2\pi k[x_j-\tau_s f(u_0(x_j))]}}{(\pi\gamma_j\tau_s k)^{1/3}}
%\int_{-\infty}^{+\infty} e^{\imath\mathrm{sgn}(\gamma_j)\frac{z^3}{3}}\ dz=
&\qquad\sim\frac{e^{\imath2\pi nx_j}e^{-\imath2\pi k[x_j-\tau_s f(u_0(x_j))]}}{(\pi\gamma_j\tau_s k)^{1/3}}\ 
\frac{2\pi}{3^{2/3}\Gamma(2/3)}\ .
\end{split}
\end{equation}
 Inserting (\ref{eq:Ij}) into (\ref{eq:uktau}) at $\tau=\tau_s$ one gets
\[
\hat u_k(\tau_s)\sim -\imath\left[\sum_{j=1}^m
\frac{u_0'(x_j)e^{-\imath2\pi k[x_j-\tau_s f(u_0(x_j))]}}{(9\pi|\gamma_j|\tau_s)^{1/3}\Gamma(2/3)}\right] k^{-4/3}\ ,
\] 
where $\Gamma(2/3)$ is the Euler gamma function at $2/3$.
Its square modulus yields the asymptotic formula 
\begin{equation}
\label{eq:fesgen}
|\hat u_k(\tau_s)|^2\sim\left|\sum_{j=1}^m\frac{u_0'(x_j)e^{-\imath2\pi k[x_j-\tau_sf(u_0(x_j))]}}{
(9\pi \tau_s \gamma_j)^{1/3}\Gamma(2/3)}\right|^2k^{-8/3} \, ,
\end{equation}
which holds as $k\to+\infty$.

Finally, formula (\ref{eq:fesgen2}) for the normalized FES is obtained by proving that the function
$d\Phi(\lambda_0(x))/dx$, which enters the definition (\ref{eq:taulr}) of the shock time
(\ref{eq:taul}), displays a single absolute maximum point $\hat x$ if $a$ is small enough. Then, formula (\ref{eq:fesgen})
simplifies to
\begin{equation}
\label{eq:laktaus}
\begin{split}
|\hat \lambda_k(\tau_s)|^2&\sim\left|\frac{u_0'(\hat x)}{
(9\pi \tau_s d^3\Phi(\lambda_0(\hat x))/dx^3)^{1/3}\Gamma(2/3)}\right|^2k^{-8/3} \\
&= C\ k^{-8/3}\ ,
\end{split}
\end{equation}
where the constant $C$ is independent of $k$. Assuming that the form of the FES at the shock time is given by 
$C k^{-8/3}$ for all $k\geq 1$, the normalized FES turns out to be
\[
\frac{E_k(t_s)}{\sum_{k>0}E_k(t_s)}=\frac{|\hat \lambda_k(\tau_s)|^2}{\sum_{k>0}|\hat \lambda_k(\tau_s)|^2}=\frac{k^{-8/3}}{\zeta_R(8/3)}\ ,
\] 
where $\zeta_R(s)=\sum_{k>0}k^{-s}$ is the Riemann zeta function. One finds numerically that
$\zeta_R(8/3)=1.28419\dots$, whose reciprocal is $0.77870\dots$, which justifies formula (\ref{eq:fesgen2}). 

\subsection{Short time evolution of the energy spectrum}\label{subsec:ShortTimes}
We now analyze the growth of the FES for short times. To this aim, we consider for simplicity the dynamics of the inviscid Burgers equation with all coefficients equal to $1$, i.e.
\begin{equation}\label{eq:Starrr}
	u_t=u u_x \, .
\end{equation}
Expanding in Fourier series, and considering only $k\geq 0$, we get
\begin{equation}\label{eq:FourierBurg}
	\frac{d \hat{u}_k}{dt} = \imath  \pi k \sum_{1 \leq p \leq k-1} \hat{u}_p \hat{u}_{k-p} + 2 \imath  \pi k \sum_{p \geq 1} \hat{u}_{k+p} \hat{u}^*_{p} \, .
\end{equation}
In order to study the initial dynamics where the high modes contain a small amount of energy, we can neglect the second term in \eqref{eq:FourierBurg}. Thus, by imposing that $\hat{u}_0=0$, we can explicitly solve the evolution equations. One immediately observes that
\[
	\begin{split}
	\frac{d \hat{u}_1}{dt}&=0  \, , \quad
	\frac{d \hat{u}_2}{dt}=\imath \pi (\hat{u}_1)^2 \, , \quad 
	\frac{d \hat{u}_3}{dt}=2 \imath \pi \hat{u}_1 \hat{u}_2 \, , \cdots
	\end{split}
\]
and then one guesses a solution of the form $\hat{u}_1(t)=\hat{u}_1(0)$, $\hat{u}_2(t) \sim t$, $\hat{u}_3(t) \sim t^2$ and so on. Thus, for the generic mode $k \in \mathbb{Z}$, one predicts
\begin{equation}\label{eq:AsymptT}
	\hat{u}_k(t) \sim t^{k-1} \, .
\end{equation}
This ansatz is compatible with \eqref{eq:FourierBurg}, since one has $\frac{d}{dt}\hat{u}_k(t)=t^{k-2}$ on the left--hand side and $t^{p-1}t^{k-p-1}=t^{k-2}$ on the right--hand side.

This analysis predicts that the initial growth of the energy of the normal modes follows the scaling law
\begin{equation}\label{eq:InitialGrowthCC}
	E_k = |\hat{u}_k|^2 \sim t^{2k-2} \, .
\end{equation}
This result can be also obtained using the explicit representation of the solution of the inviscid Burgers equation and the integral representation of the Bessel functions (see Appendix \ref{sec:AppA}). However, solving the equations using the iterative procedure described above, describes the physical origin of the scaling, which comes from an energy cascade from longwavelength modes to the shorter ones. Indeed, in neglecting the second term on the right--hand side of \eqref{eq:FourierBurg}, we are neglecting the effects of a backward cascade of energy and we are keeping only the forward cascade. 
%{\Large \color{red} \texttt{STEFANO: QUALCHE COMMENTO PIU' FISICO?}}

In Fig.\ \ref{fig:Growth} we interpolate the initial growth of the energies of the first modes with straight lines and we see a general agreement with eq. \eqref{eq:InitialGrowthCC}. Thus, we can conclude that the forward energy cascade is the physical phenomenon underlying the formation of the power--law scaling in the energy spectrum.

\begin{figure}[t]
	\includegraphics[width=0.46\textwidth]{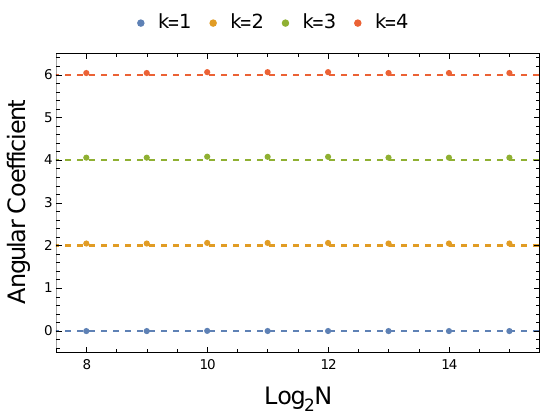}
	
	\caption{Angular coefficients of lines interpolating the growth of Fourier modes (see Fig. \ref{fig:Growth}) with indices $k=1,2,3,4$ for different values of $N$ showing the persistence of the slopes when increasing $N$ with fixed $\varepsilon$. The initial datum is a traveling wave excitation with $\alpha=1$, $\beta=\frac{1}{2}$ and $\varepsilon=0.0005$.}\label{fig:AngularN}
\end{figure}

We present in Fig.\ \ref{fig:AngularN} the angular coefficients of the energies of the normal modes with $k=1,\dots,4$ when the number of particles increases. This shows that the scenario described by the Burgers equation is stable in the thermodynamic limit.

For initial data that are not exactly TWEs, one observes that the behavior of Fig.\ \ref{fig:Growth} remains the same, apart from the additional presence of small amplitude oscillations around the interpolating line.  

Differently from the asymptotics \eqref{eq:fesgen2}, formula \eqref{eq:AsymptT} strongly relies on the form of the nonlinearity. Therefore, when the energy of the initial datum is increased, the monomial with higher degree in \eqref{eq:larhoev} cannot be neglected anymore and this effect dramatically changes the exponent of the power--law of the time evolution.

When the energy of the initial datum increases, the presence of a non-zero counterpropagating traveling wave starts to modify the short-time behavior. This phenomenon is visibile e.g. in Fig.\ 2 in \cite{PCSF11}. Indeed, in that case, one cannot neglect either the contribution of the field $\rho$ or the correction of the initial datum given by the canonical transformation. In fact, for higher energies, one observes a more complex law describing the growth of energies of normal modes. 

\subsection{Time evolution of the Fourier energy spectrum: From turbulence to equipartition}\label{subsec:AfterTS}

\begin{figure}[h!]
	\includegraphics[width=0.43\textwidth]{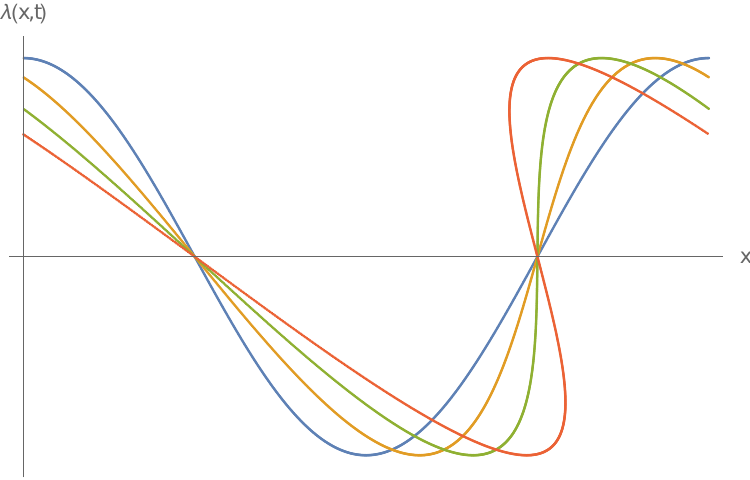}
	
	\caption{Solution of the Burgers equation using the method of characteristics with cosine initial datum. Notice that, after the shock time, the solution consists of a multi-valued function.} \label{fig:MultiVal}
\end{figure}

A solution of the generalized Burgers equations $u_\tau=f(u) u_x$  in terms of a function no longer exists for times $\tau>\tau_s=t_s/N$. Nevertheless, it is possible to give a geometric meaning to the solution of the generalized Burgers equations also for larger times by means of the method of the characteristics. After the shock time, this \emph{geometric solution} consists in a multi-valued function (see Fig.\ \ref{fig:MultiVal}).

It is clear that, from a physical point of view, a multi-valued solution can not be the correct description of the time evolution for the lattice FPUT model since the velocity and the position of each particle are determined uniquely by the solution of the equations of motion. 

For the FPUT system, what could be called the ``gradient catastrophe'', which occurs at the shock time, implies a transfer of energy to the highest Fourier modes of wavelength $\sim 1/N$. This effect doesn't allow us to take a continuum limit after the shock over the whole spatial domain following the procedure outlined in Sec.~\ref{subsec:ContinuousModel}. In order to obtain a correct continuum description of the shock region, higher order derivatives
of the fields must be taken into account. This procedure amounts to replacing the Burgers equations with a pair of KdV  equations~\cite{GP22,PB05,BP06,GPR}. However, away from the shock spatial region, the Burgers equation can still be used to describe the FPUT dynamics. The idea of describing the dynamics after the shock using the Burgers equation is quite intuitive. Nonetheless, it is a challenging mathematical problem to relate the solution of the KdV equation to its dispersionless limit, the Burgers equation. A very recent and pioneering result, in this direction, has been only recently obtained in the study of the Benjamin-Ono equation in \cite{Gerard-BO}. In this case, the zero-dispersion limit of the solution of the Benjamin-Ono equation is given by a weighted mean of the values of the multi-valued solution of the Burgers equation. On one hand, this approach gives a rigorous connection between the geometric solution of the Burgers equation and the zero-dispersion limit of the Benjamin-Ono, on the other hand  it justifies the heuristic idea that, far from the shock, where the geometric solution is given by a one-valued function, the analytical solution of the Burgers equation still has meaning.  Nevertheless, it is not clear whether that analysis can be adapted to the KdV equation.

Using the idea that, far from the shock, the analytical solution of the Burgers equation still has a meaning, for $\tau>\tau_s$ we will proceed heuristically. Indeed, let us consider the left TWE $\lambda(x,0)=2\cos(2\pi x)$ with $\beta=1/2$ in order to eliminate the quadratic term in $a$ in eq.~\eqref{eq:Phi}. In this case system \eqref{eq:larhoev} yields the Burgers equation
$\lambda_\tau=(a\alpha/\sqrt{2})\lambda\lambda_x$, whose solution is obtained from the implicit equation $\lambda=2\cos\left(2\pi(x+(a\alpha/\sqrt{2})\lambda \tau)\right)$.
The cosine initial condition is progressively deformed into a sawtooth profile $\sigma(x)$ with the discontinuity at $x=3/4$ (the point in which the initial cosine vanishes and the profile has positive derivative) and slope $-4$. Performing a Fourier transform, one finds that
\begin{equation} 
\label{eq:saw}
\sigma(x)=\sum_{k\neq0}\frac{2}{\imath\pi k}e^{\imath2\pi k(x+1/4)}\ .
\end{equation}
Using the method of characteristics, it is possible to see that the time needed for the position of the maximum of the initial cosine to reach the node at $x=3/4$ is $(\pi/2)\tau_s$, thus larger than the shock time $\tau_s$.  At the shock time $\tau_s$, the spatial derivative of $\lambda$ becomes infinite in the Burgers equation, huge but finite on the lattice due to dispersion. The formation of the sawtooth profile then follows in time
the creation of the shock. Therefore, after the formation of the sawtooth profile, we can decompose the wave profile as $\lambda(x,\tau)=\sigma(x)+r(x,\tau)$, where the deviation $r$ with respect to the sawtooth profile \eqref{eq:saw} is a smooth function. The Fourier coefficients of $\sigma(x)$ decay 
as $1/k$, while those of the smooth deviation $r$ can be shown to decay faster~\cite{cinfinity}.  Therefore, the FES of 
$\lambda$ is dominated by $|\hat\sigma_k|^2\propto k^{-2}$.  This heuristic argument can be verified in numerical experiments by measuring 
the slope of the FES after the shock time $\tau_s$. The time evolution of the slope is shown in Fig.~\ref{fig:ZetaT}: one observes an
extended time window, which depends on the specific energy of the initial condition, where $\zeta=2$. For example, in the case reported in Fig.~\ref{fig:ZetaT} with $\varepsilon=0.005$ the time window is approximately from two to six shock times. The exponent $\zeta=2$  was first found for Burgers turbulence in \cite{OriginalBurgers} and its relevance was already established in~\cite{FF} and further analyzed in~\cite{Kida}. From a mathematical point of view, the exponent $\zeta=2$ in the Burgers spectrum has been established in the zero-viscosity limit for the viscous Burgers equation, see the recent review \cite{KuksinRev}. 

At later times, although the numerical determination of the slope becomes much harder, Fig.~\ref{fig:ZetaT} suggests that the slope eventually increases, detecting a trend to equipartition, which corresponds to a vanishing slope and the disappearance of the exponential fall off (see Fig.~\ref{fig:Spectra}).

\section{Conclusions}
In this paper we have highlighted the presence of a power--law scaling region in the spectrum of the Fermi-Pasta-Ulam-Tsingou chain. To the best of our knowledge, previous works never detected such a region and concentrated on the exponential cut-off which is present for short wavelengths. A previous discussion of this scaling region can be found in \cite{PRL}. We have discussed in detail the value of the scaling exponents and we have found that its phenomenology is well interpreted by mapping the FPUT dynamics onto the Burgers equation performing the thermodynamic limit in an appropriate way. The mapping to the Burgers equation also allows us to determine the early time growth of the energy of each mode, which is again a power--law in time with a mode-dependent exponent which can be determined analitically. This analytical calculation unveils the underlying physical mechanism behind the scaling, i.e. an energy cascade from the long wavelength modes to shorter ones.

We have derived the Burgers equation by defining renormalized fields $\lambda$ and $\rho$ (see Sec.~\ref{sec:PerturbationTheory}) that are dressed versions of the left and right traveling waves of the linear model in the thermodynamic limit. Using the dynamics expressed in terms of these fields, we can derive the time evolution of the energy of the Fourier modes $E_k(t)$ for short times. We find a precise scaling law describing the energy gain of modes $k\geq 2$ (see Sec.~\ref{subsec:ShortTimes}). The approximate equation for the mode amplitudes, when solved recursively, gives an insight into the physical phenomenon underlying the formation of the so-called ``metastable packet'' \cite{Gallavotti-FPU-Libro}. Another important prediction obtained using the Burgers equation is the power--law spectrum $E_k(t_s) \sim k^{-8/3}$ at the shock time, see Sec.~\ref{subsec:ShockTime}.

At later times, the solution of the Burgers equation is described by a multivalued function and, therefore, cannot be considered as the thermodynamic limit of the FPUT lattice model. Nevertheless, at such later times, one observes numerically a large portion of the spectrum with $E_k \sim k^{-2}$. Again, the presence of this further power--law, can be related to the Burgers equation \cite{OriginalBurgers,KuksinRev,Kida,NewBurg}. This exponent has not yet been derived analytically for the Burgers equation or for the lattice dynamics and will be the subject of further investigations.

Our paper opens the way to a more careful investigation of the power--spectrum of the FPUT dynamics revealing a power--law scaling. Although mainly focused on one-mode initial conditions, some of the results discussed in this paper extend to a broader class of initial conditions. Specifically, the theorem in Sec.~\ref{subsec:ShockTime} shows that when the initial data consist of a finite (but arbitrarily large) linear combination of Fourier modes, the FES at the shock time exhibits the asymptotic power--law $E_k \sim k^{-8/3}$. Additionally, it is known from~\cite{Kida} that the solution to the Burgers equation exhibits a power--law spectrum $E_k \sim k^{-2}$ for a very general set of initial data. Beyond this, the behavior of the shock time and the precise width of the aforementioned spectral windows may reveal intriguing and non-trivial features when a larger number of modes is excited. In particular, in this latter case, careful consideration of the role of relative phases is crucial (see~\cite{BLP09}), and we will dedicate a future paper to explore this problem. Preliminary results are available in \cite{Druais}.

\appendix

\section{Initial growth}\label{sec:AppA}
\begin{figure}[t]
	\includegraphics[width=0.46\textwidth]{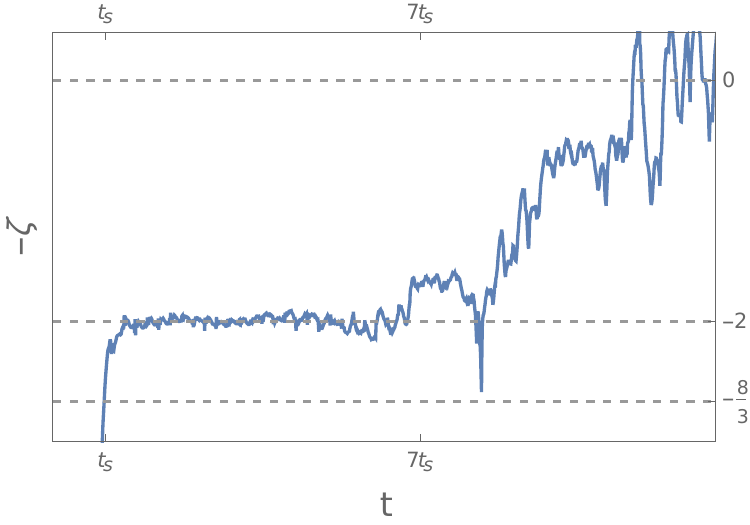}
	
	\caption{$\zeta$ exponent, see \eqref{eq:DefZeta}, of the power--law region of the spectrum of the FPUT dynamics as a function of time measured in units of the shock time (see \eqref{eq:taulr}) of the Burgers equation. We observe an initial growth of $\zeta$ with a value $8/3$ at the shock time. Around twice the value of the shock time, $\zeta$ reaches the value $2$ which persists for about six shock times. A later trend to equipartition $\zeta=0$ is also observed. The initial datum is a traveling wave excitation with $\varepsilon=0.005$, $\alpha=1$ and $\beta=0.5$ $N=8192$.}\label{fig:ZetaT}
\end{figure}

We can write the $k$th Fourier component of the solution of \eqref{eq:Starrr} using Eq.~(9.1.21) in \cite{AS} as a representation for the Bessel function of the first kind $J$ (see (9.1.10) in \cite{AS}):
\begin{equation} \label{eq:SolutionBessel}
	\widehat{u}_k(t) \;=\; \frac{ \mathrm{i} e^{- \mathrm{i} k \frac{\pi}{2}}}{2 \pi k t} J_k(2 \pi k A t) \, .
\end{equation}
With this last expression at hand, we get
\begin{equation}\label{eq:SpecificEnergy}
	E_k(t)  \;=\; \frac{J_k^2(2 \pi k A t)}{4 \pi^2 k^2 t^2} \,.
\end{equation}
To obtain a theoretical prediction on the growth of the energy of the normal modes for short times, we use the asymptotic expansion of the Bessel function for small argument (eq. (9.1.19) in \cite{AS}) to obtain
\begin{equation}\label{eq:EnergiesShortTime}
%	E_k(t) \sim \frac{1}{8} \frac{\pi^{2k-3} e^{2k} A^{2k}}{k^3} t^{2k -2} \, , \quad t \ll 1 \, .
E_k(t) \sim t^{2k-2} \, \quad t \ll 1 \, .
\end{equation}

\section{Rigorous Mathematical Results}
The FPUT problem had a dramatic impact on the mathematical literature. Without aiming at a complete overview, that would require an indipendent review by itself, let us quote the main mathematical results that led to the present understanding of the phenomenon. 

\subsection{Connection with KdV}
Historically, the attempt to explain the FPUT paraox is one of the motivations to analyze the dynamics of the Korteweg--de Vries equation by Zabusky and Kruskal \cite{ZK65}. In that paper the authors analyze numerically the solution of the Korteweg-de Vries equation on the interval with periodic boundary conditions. They give a phenomenological description of some aspects of the near-recurrence of the initial datum in terms of the interaction of solitons.

In few years it was then discovered that Korteweg--de Vries is an integrable infinite dymensional system \cite{Kdv1,Kdv2,Kdv3,Kdv4}. This was the first connection bewteen the behavior of FPUT chain in the metastable state and an underlying integrable dynamics.

In the early '2000s a renewed interest in the connection between FPUT and KdV permits to prove rigorously the existence of the metastability phenomenon \cite{BP06}.  In this paper, the authors show that the resonant normal form of the FPUT lattice consists in a pair of counterpropagating KdV equations. Then, a further step of normal form is performed in \cite{GP22} where it is shown that the standard methods of Hamiltonian perturbation theory are not enough to guarantee integrability of the second order normal form. Nevertheless, a wider class of transformation guarantees integrability at second order explaining at a formal level the persistence of the metastable packet for time--scales comparable with those observed in numerical experiments. Last, the construction of Zabusky and Kruskal is reveiwed in \cite{GPR} where it is shown that, restricting to almost one directional waves, in general the third order normal form is \emph{not} integrable. Integrability is recovered by choosing special parameters, such as the parameters corresponding to the series expansion of the Toda lattice (see below).

\subsection{Connection with Toda}
The first connection between the FPUT lattice and the Toda chain, to the present state of reconstruction, is due to Manakov \cite{Man}. The Hamiltonian of the Toda chain has the form \eqref{eq:H}, where $V(z)$ is replaced by
\begin{equation}
	V_{\text{Toda}}(z)=\frac{1}{4 \alpha^2} \big(e^{2 \alpha z}-1 - 2 \alpha z \big) \, .
\end{equation}
Expanding $V_{\text{Toda}}(z)$ for $z$ small, one observes that 
\begin{equation}
	V(z)-V_{\text{Toda}}(z)=\frac{1}{4}\left(\beta-\frac{2 \alpha^2}{3}\right)z^4 + O(z^5)
\end{equation}
and, therefore, for small values of specific energies, the Toda model is an integrable model closer to the FPUT than the harmonic chain. With respect to KdV, Toda permits an approximate analysis without requiring the hypothesis of long--wavelength initial excitation. 

The connection between Toda and FPUT was re-proposed in \cite{FFML} where, on the one side, it is shown that the dynamics of the Toda chain is practically undistinguishable from the dynamics of the FPUT for a certain time--scale (compare e.g. Fis. 25 and 26 in \cite{FFML}) and, on the other side, a way to compute numerically the actions of the Toda chain is provided. 

It is impossible to give a complete overview on the connection between Toda and FPUT, for a more detailed discussion on the connection between the two models we refer the reader to \cite{BPCerci} and references therein.

Let us stress that, beside the change of paradigm introduced by discovering the vicinity of Toda and FPUT, the connection between Toda actions and harmonic normal modes has been rigorously established only in the (unphysical) regime of parameters $E \ll N^4$ \cite{BambusiMaspero}. The map between Toda integrals of motion and the intial datum is less explicit than the relation between FPUT and KdV.

\subsection{KAM-like results}
Recurrences similar to those observed in the FPUT original experiment, are typical of integrable systems. It is therefore natural to ask whether a Kolmogorov-Arnold-Moser (KAM) theorem could be applied to the FPUT Hamiltonian. The procedure is not trivial since one of the hypotheses of the KAM theorem is the non-degeneracy of the unperturbed Hamiltonian, that is: when considering a Hamiltonian system written in action-angle variables of $H_0$, $H(I,\varphi)=H_{0}(I)+P(I,\varphi)$, one must require $\det \frac{\partial^2 H_{0}(I)}{\partial I \partial I} \neq 0$. This is manifestly violated when one chooses as an unperturbed model the chain of harmonic oscillators, whose Hamiltonian in action-angle variables is $H_0(I)=\sum_{k} \omega_k I_k$. To overcome this problem, one could perform one step of the Birkhoff normal form and hope that, after this transformation, the transformed Hamiltonian can be split ino an integrable part $H_0$ which satisfies the non--degeneracy condition and a smaller perturbation.

The first result obtained following this strategy is due to Nishida \cite{Nishida} who proved that, for $\alpha=0$ and under very particular conditions on the number of particles in the chain, the KAM theorem was applicable. The hypothesis on the number of particles, was then relaxed by Rink \cite{Rink} in a remarkable paper where he shows that after one step of the Birkhoff normal form, the Hamiltonian of the $\beta$--model satisfies the hypothesis to apply the KAM theorem.

Later, in a pair of works, Henrici and Kappeler \cite{HenriciKappeler,KappelerHenrici} proved that the same result holds also for the $\alpha+\beta$ chain. In this case, part of the difficulty is introduced by the fact that the $\alpha+\beta$ models have less symmetries than the pure $\beta$ model.  

The physical implication of the KAM theorem is that if we fix the number of particles $N$, there exists a treshold $E_*$, depending on $N$, such that if the energy of the initial datum is less or equal to $E_*$ then the surface $H(q,p)=E$ admits a decomposition into two disjoint set of strictly positive measure that are invariant under the flow of the Hamiltonian and then, by Birkhoff theorem, the system is \emph{not} ergodic (and hence, it never thermalizes in the Gibbs sense).

\begin{acknowledgments}
This work was partially supported by GNFM (INdAM) the Italian national group for Mathematical Physics. M.G. acknowledges financial support from the European Research Council (ERC) under the European Union’s Horizon 2020 research and innovation program ERC StG MaMBoQ, n.802901. S.R. acknowledges financial support from the MUR- PRIN2017 project ``Coarse-grained description for non- equilibrium systems and transport phenomena (CONEST)'' No. 201798CZL.
\end{acknowledgments}

\nocite{*}
\providecommand{\noopsort}[1]{}\providecommand{\singleletter}[1]{#1}%


\begin{thebibliography}{10}

%\bibitem{suppmat} \emph{Supplemental Material} with all details of the computations. 

\bibitem{Kinchin} A. Y. Kinchin, \emph{Mathematical foundations of statistical mechanics} (Dover, New York, 1949).

\bibitem{CerciBook} C. Cercignani, \emph{The Boltzmann Equation and Its Applications} (Springer, New York, NY, 1988).

\bibitem{Ashcroft-Mermin} N. W. Ashcroft, N. D. Mermin, \emph{Solid State Physics} (Saunders College Publishing, New York, 1976).

\bibitem{FPU55}
E.~Fermi, J.~Pasta, and S.~Ulam, Studies of nonlinear problems, Los-Alamos Internal Report, Document LA-1940 (Los Alamos National Laboratory, Los Alamos, 1955).

\bibitem{Daux08}
T.~Dauxois, Fermi, Pasta, Ulam and a mysterious lady, Phys. Today P{\bf61}, 55 (2008).

\bibitem{Gallavotti-FPU-Libro} G. Gallavotti, \emph{The Fermi-Pasta-Ulam Problem: A Status Report}, (Springer-Verlag, Berlin, Heidelberg, 2008).

\bibitem{Chaos-VolumeCollettivo} G.~P.~Berman, F.~Izrailev, The Fermi–Pasta–Ulam problem: fifty years of progress,  Chaos \textbf{15}, 015104 (2005).

\bibitem{Kino} T.~Kinoshita, T.~Wenger and D.S.~Weiss, A quantum Newton's cradle, Nature (London) {\bf440},
900 (2006).


\bibitem{Rigol-PRL} A. D. Ribeiro, M. Novaes, and M. A. de Aguiar, Uniform Approximation for the coherent-state propagator using a conjugate application of the Bargmann representation, Phys. Rev. Lett {\bf95}, 050405 (2005).

\bibitem{RevFloq} W. W. Ho, T. Mori, D. A. Abanin, E. G. Dalla Torre, Quantum and classical Floquet prethermalization, Ann. Phys. {\bf454}, 169297 (2023).

\bibitem{Abanin} D. Abanin, W. De Roeck, W. W. Ho, F. Huveneers, A rigorous theory of many-body prethermalization for periodically driven and closed quantum systems, Commun. Math. Phys. {\bf354}, 809 (2017).

\bibitem{Else} E. V. Else, W. W. Ho, P. T. Dumitrescu, Long-lived interacting phases of matter protected by multiple time-translation symmetries in quasiperiodicaly driven systems, Phys. Rev. X {\bf10}, 021032 (2020).

\bibitem{GL} M. Gallone, B. Langella, Prethermalization and conservations laws in quasi-periodically driven quantum systems, J. Stat. Phys. {\bf191}, 100 (2024).

\bibitem{TodaIntegrable} M. H\'enon, Integrals of the Toda lattices,
Phys. Rev. B {\bf9}, 1921 (1974)

\bibitem{Po03}
A.~Ponno, Soliton theory and the Fermi-Pasta-Ulam problem in the thermodynamic limit, Europhys. Lett. {\bf64}, 606 (2003).

\bibitem{BP11} G. Benettin and A. Ponno, Time-Scales to equipartition in the Fermi-Pasta-Ulam problem: finite size effects and thermodynamic limiti, J. Stat. Phys. {\bf144}, 793 (2011).

\bibitem{IC66} F.M.~Izrailev and B.V.~Chirikov, Statistical properties of a nonlinear string, Sov. Phys. Dokl. {\bf11}, 30 (1966).

\bibitem{Izrailev2} B. V. Chirikov, F. M. Izrailev, and V. A. Tayurski, Numerical experiments on the statistical behaviour of dynamical systems with a few degrees of freedom, Comput. Phys. Commun. {\bf5}, 11 (1973).

\bibitem{PRARuffo} R. Livi, M. Pettini, S. Ruffo, M. Sparpaglione, and A. Vulpiani, Equipartition threshold in nonlinear large Hamiltonian systems: The Fermi-Pasta-Ulam model, Phys. Rev. A {\bf31}, 1039 – Published 1 February 1985

\bibitem{JPA-Livi-Politi-Ruffo} R. Livi, A. Politi, and S. Ruffo, Scaling-law for the maximal Lyapunov exponent, J. Phys. A: Math. Gen.  {\bf25} 4813 (1992).

\bibitem{PRL} M.~Gallone, M.~Marian, A.~Ponno, S.~Ruffo, Burgers turbulence in the Fermi-Pasta-Ulam-Tsingou chain, Phys. Rev. Lett. {\bf 129} 114101 (2022).

\bibitem{FF}
J.D.~Fournier and U.~Frisch, L'\'equation deBurgers d\'eterministe et statistique, J. Mec. Theor. Appl. {\bf2}, 699 (1983).

\bibitem{Man} S.V. Manakov, Complete integrability and stochastization of discrete dynamical systems, Sov. Phys. JEPT {\bf40}, 269 (1974).

\bibitem{Hofstrand} A. Hofstrand, Near-integrable dynamics of the Fermi-Pasta-Ulam-Tsingou problem, Phys. Rev. E \textbf{109} 034204 (2024).

\bibitem{DeNardis2024} L. Biagetti, C. Guillaume, J. De Nardis, Three-stage thermalization of a quasi-integrable system, Phys. Rev. Res. \textbf{6} 023083 (2024).

\bibitem{ZK65}
N.J.~Zabusky and M.D.~Kruskal, Interaction of ``Solitons'' in a collisionless plasma and the recurrence of initial states, Phys. Rev. Lett. {\bf15}, 240 (1965).

\bibitem{GP22} M. Gallone and A. Ponno, \emph{Hamiltonian field theory close to non dispersive strings}, in \emph{Qualitative Properties of Dispersive PDEs}, edited by V. Georgiev, A. Michelangeli, R. Scandone, Springer INdAM Series, vol 52. (Springer, Singapore, 2021).


\bibitem{Ref1} I.M. Gelfand and S.V. Fomin, \emph{Calculus of Variations} (Dover, New York, 2000).


\bibitem{Ref2}  J. E. Marsden and T.S. Ratiu, \emph{Introduction to Mechanics and Symmetry}, 2nd ed. (Springer, New York, 1999).

\bibitem{Giorgilli} A. Giorgilli, \emph{Notes on Hamiltonian Dynamical Systems} (Cambridge University Press, Camrbidge, UK, 2022).


\bibitem{Ref3} A.~Erd\'elyi, \emph{Asymptotic Expansions} (Dover, New York, 1956).

\bibitem{PCSF11}
A.~Ponno, H.~Christodoulidi, Ch.~Skokos, and S.~Flach, The two-stage dynamics in the Fermi-Pasta-Ulam problem: from regular to diffusive behavior, Chaos {\bf21}, 043127 (2011).

\bibitem{PB05}
A.~Ponno and D.~Bambusi, Korteweg-de Vries equation and energy sharing in Fermi-Pasta-Ulam, Chaos {\bf15}, 015107 (2005).


\bibitem{BP06} D. Bambusi and A. Ponno, On metastability in FPU, Commun. Math. Phys. {\bf264}, 539 (2006).

\bibitem{GPR}
M.~Gallone, A.~Ponno, and B.~Rink, Korteweg-de Vries and Fermi-Pasta-Ulam-Tsingou: asymptotic integrability of quasi unidirectional waves, J. Phys. A: Math. Theor. {\bf54}, 305701 (2021).

\bibitem{Gerard-BO} P. G\'erard, The zero dispersion limit for the Benjamin-Ono equation on the line, C. R. Math. {\bf362} 619 (2024).

\bibitem{cinfinity} If $f(x)$ is a smooth function, then its fourier components satisfy $|\hat{f}_k| \leq C_N (1+|k|)^{-N}$ for any $N>0$.

\bibitem{OriginalBurgers} J.M. Burgers, A mathematical model illustrating the theory of turbulence, Adv. Appl. Mech. {\bf 1} 171 (1948).

\bibitem{Kida} S. Kida, Asymptotic properties of Burgers turbulence, J. Fluid Mech. {\bf 93}, 337 (1979).

\bibitem{KuksinRev} S. Kuksin, The K41 theory and turbulence in 1D Burgers equation, Chaos {\bf34}, 022103 (2024).

\bibitem{NewBurg} A. Das, P. Dutta, and V. Shukla, Poles, shocks, and tygers: The time-reversible Burgers equation,
Phys. Rev. E {\bf109}, 065108 (2024).

\bibitem{BLP09} G. Benettin, R. Livi and A. Ponno, The Fermi-Pasta-Ulam problem: scaling laws vs. initial conditions, J. Stat. Phys. {\bf135}, 873 (2009).

\bibitem{Druais} E. Druais, Internship report

\bibitem{AS} M. Abramowitz and I.A. Stegun, \emph{Handbook of Mathematical Functions}, (Dover, New York, 1965).

\bibitem{Kdv1} C. S. Gardner, J. M. Greene, M. D. Kruskal, and R. M. Miura, Method for solving the Korteweg-de Vries equation,
Phys. Rev. Lett. {\bf19}, 1095 (1967). 

\bibitem{Kdv2} P.D. Lax, Integrals of Nonlinear Equations of Evolution and Solitary Waves, Comm. Pure
Appl. Math. {\bf21}, 467 (1968).

\bibitem{Kdv3} R.M Miura, C.S. Gardner and M.D. Kruskal, Korteweg-de Vries equation and generalization,
II. Existence of conservation laws and constants of motion, J. Math. Phys. {\bf9}, 1204 (1968).

\bibitem{Kdv4} V.E. Zakharov and L.D. Feddeev, Korteweg-de Vries Equation: a completely integrable Hamil-
tonian system, Funct. Analysis Appl. {\bf5}, 280 (1971).

\bibitem{FFML}
W.~Ferguson, H.~Flaschka and D.~McLaughlin, Nonlinear normal modes for the Toda Chain, J. Comput. Phys. {\bf45}, 157 (1982).

\bibitem{BPCerci} G. Benettin, A. Ponno,  FPU model and Toda model: A survey, a view, in \emph{From Kinetic Theory to Turbulence Modeling}, edited by P. Barbante, F.D. Belgiorno, S. Lorenzani, L. Valdettaro, Springer INdAM Series, vol 51. (Springer, Singapore, 2023).

\bibitem{BambusiMaspero} D. Bambusi, A. Maspero, Birkhoff coordinates for the Toda Lattice in the limit of infinitely many particles with an application to FPU, J. Funct. Anal. {\bf270}, 1818 (2016).

\bibitem{Nishida} T. Nishida, A note on an existence of conditionally periodic oscillation in a one-dimensional anharmonic
lattice. Mem. Fac. Eng. Univ. Kyoto {\bf33}, 27 (1971).

\bibitem{Rink} B. Rink, Symmetry and Resonance in Periodic FPU Chains, Commun. Math. Phys. {\bf218}, 665 (2001).

\bibitem{HenriciKappeler} A. Henrici,T. Kappeler, Results on normal forms for FPU chains, Commun. Math. Phys. {\bf278}, 145 (2008).

\bibitem{KappelerHenrici} T. Kappeler, A. Henrici, Resonant normal form for even periodic FPU chains, J. Eur. Math. Soc. {\bf11} 1025 (2009).



\end{thebibliography}
\end{document}